\documentclass[a4paper,12pt]{article}
\usepackage{times,mathptm,amsmath,amssymb,mathrsfs}
\usepackage{graphicx,here,cite,fancyhdr,enumerate}
\usepackage[margin=25mm]{geometry}
\newcommand{\nn}{\nonumber\\}
\newcommand{\TT}{{\cal T}}

\newcommand{\bPsi}{\Psi^\dagger}
\newcommand{\tV}{\tilde{V}}
\newcommand{\tv}{\tilde{v}}

\newcommand{\tT}{\tilde{T}}
\newcommand{\tD}{\tilde{D}}

\newcommand{\tG}{\tilde{G}}
\newcommand{\be}{\begin{equation}}
\newcommand{\ee}{\end{equation}}
\newcommand{\bn}{\bar{n}}
\newcommand{\w}{\omega}
\newcommand{\p}{{\bf p}}
\newcommand{\la}{\langle}
\newcommand{\ra}{\rangle}

\newcommand{\eqn}[1]{\label{#1}}
\newcommand{\eq}[1]{Eq.~(\ref{#1})}

\makeatletter
\renewcommand{\section}{\@startsection{section}{1}{\z@}%
                                     {-3.25ex\@plus -1ex \@minus -.2ex}%
                                     {1.5ex \@plus .2ex}%
                                     {\textnormal\normalsize\bfseries}}
\renewcommand{\subsection}{\@startsection{subsection}{2}{\z@}%
                                     {-3.25ex\@plus -1ex \@minus -.2ex}%
                                     {1.5ex \@plus .2ex}%
                                     {\textnormal\normalsize\bfseries}}
\makeatother
\begin{document}
\fancyhead[L]{\textit{Topic area:}~\textbf{NUPP}}
\fancyfoot[L]{\textit{Presenting author's name:}\hspace*{2em}%
\textbf{B.~Blankleider}}
\begin{raggedright}
\section*{\large In-matter three-body problem}
A.N.~Kvinikhidze$^{1}$ and \underline{B.~Blankleider}$^{2}$\\
\vspace*{1ex}\ \\
$^{1}$\textit{The Mathematical 
Institute of Georgian Academy of Sciences, Tbilisi, Georgia, }\\
$^{2}$\textit{School of Chemistry, Physics, and Earth Sciences, Flinders University, Beford Park, SA 5042, Australia}\\
\vspace*{0.5ex}\ \\

\textit{e-mail of corresponding author:}  boris.blankleider@flinders.edu.au
%
%

\subsection*{Introduction} 
The in-matter three-body problem plays an important role in describing a large variety of interesting phenomena in many-body systems. For example, in order to understand the formation of bound states in heavy ion collisions, three-body calculations are needed to study the modification of
the binding energy and wave function of a three-nucleon bound state due to nuclear matter
of finite density and temperature \cite{Beyer1,Beyer2}. Similarly, studies of the binding energy of three quarks are of relevance to the understanding of color superconductivity and phase transitions in quark matter \cite{Beyer_quark,Mattiello}.

Although hole contributions make even non-relativistic in-matter descriptions four-dimensional, the numerical complications of a four-dimensional approach makes a three-dimensional reduction desirable.  In the early work of Schuck, Villars, and Ring \cite{schuck}, equal-time Green functions were used to obtain a three-dimensional field theoretic description. To derive their equation for the zero-temperature equal time three-body wave function, they approximated the effective pair-interaction kernels by terms linear in the physical two-body potentials. 
Since the exact expression for the effective pair-interaction kernel involves an infinite series of higher order terms as well [see \eq{tVipert}], the linear approximation cannot be considered as satisfactory for the strong coupling case, e.g., when two-body bound states are possible. Despite this, the model of Ref.\ \cite{schuck} has remained to the present day the state-of-the art formulation and forms the basis of recent calculations \cite{Beyer1,Beyer2,Beyer_quark,Mattiello,Beyer3,Beyer4,BR}.

The goal of the present paper is to formulate three-dimensional equations for the finite temperature in-matter three-body problem, that take into account the full infinite series for the effective pair-interaction kernel, so that  {\em all} possible two-body sub-processes allowed by the underlying Hamiltonian are retained.

\subsection*{In-matter four-dimensional three-body equations}

The interactions of three identical particles at finite density and temperature
are described in quantum field theory by the Green function ${\cal G}$ defined by
\begin{eqnarray}\label{G6pt}
\lefteqn{(2\pi)^4\delta^4(p'_1+p'_2+p'_3-p_1-p_2-p_3){\cal G}(p'_1p'_2p'_3;p_1p_2p_3)= 
\int d^4y_1d^4y_2d^4y_3d^4x_1d^4x_2d^4x_3} 
\nn
&&e^{i(p'_1\cdot y_1+p'_2\cdot y_2+p'_3\cdot y_3-p_1\cdot x_1-p_2\cdot x_2-p_3\cdot x_3)}\,
\mbox{Tr}\left\{ \rho\, \TT [\Psi(y_1)\Psi(y_2)\Psi(y_3)
\bPsi(x_3)\bPsi(x_2)\bPsi(x_1)]\right\}\hspace{6mm}
\end{eqnarray}
where $\Psi$ and $\bPsi$ are Heisenberg fields, $\TT$ is the time ordering
operator and
$
\rho = e^{-\beta (H-\mu N)}/\mbox{Tr}\, e^{-\beta (H-\mu N)} .
$
is the statistical operator of the grand canonical ensemble \cite{walecka}. Besides being the central quantity for the description of three-body observables, this Green function is also needed
to calculate the vacuum properties of the system with the help of the dressed
single particle propagator; for example, in the four-point interaction model,
the single particle self-energy diagram is completely defined by particle-particle-hole (pph)
Green function \cite{dukelsky}.

For non-zero temperature,
two types of perturbation theory, so-called "imaginary-time"  and "real-time",
have been derived for \eq{G6pt} \cite{LeBellac}.
Here we shall use the real-time formulation of perturbation theory in which the number of degrees is doubled \cite{LeBellac,smilga}; this complication, with respect to the zero-temperature case, comes from the sum over the complete set of states  (trace) in \eq{G6pt}
(a similar discussion on the basis of the imaginary-time formalism will be presented elsewhere).
For example, the nonrelativistic free one-body propagator is given by \cite{umezawa}
\be\label{df}
d^f(p)=i\left[\frac{\bn(\p)}{p^0-\w+ i \epsilon} +
\frac{n(\p)}{p^0-\w- i \epsilon} \right]
\ee
where $\w=\w_\p=\p^2/2m - \mu$, $\mu$ is the chemical potential, and $n, \bar n$ are $2\times 2$ matrices whose elements are functions of the distribution function
$f(\w) = (e^{\beta\w}\pm 1)^{-1},$
where the upper sign ($+$) is for fermions and the lower sign ($-$) is for bosons.  The $n$ and $\bn$ satisfy the following relations which define their projection properties: $n + \bn = g$, 
$n\, g\, \bn = \bn\, g\, n = 0$,  $n \,g \,n = n$, and  $\bn\, g\,\bn = \bn$, where $g$ is a $2\times 2$ matrix with elements $g_{11}=1$, $g_{12}=g_{21}=0$, and $g_{22}=\pm 1$.
Correspondingly, elementary vertices have an extra double-valued index for 
each particle leg.

For the identical particle case considered here,  the field theoretic expression of \eq{G6pt} automatically guarantees the appropriate symmetry of the three-particle Green function ${\cal G}$.  Moreover, in the doubled degrees of freedom formalism, the matrix Green function $G$ whose first diagonal element is ${\cal G}$, is likewise properly symmetric in the case of bosons, and antisymmetric in the case of fermions. One can write \cite{kbgaug}
\be
G=G_0^P + G_0VG  \eqn{Gas}
\ee
where $G_0$ is the product of single particle dressed propagators $d(p_i)$,
\be
G_0(p_1' p_2' p_3',p_1p_2p_3)=d(p_1)\, d(p_2)\, d(p_3)\, (2\pi)^4 \delta(p_2'-p_2)\,
(2\pi)^4\delta(p_3'-p_3) ,   \eqn{G_0123}
\ee
$G_0^P$ is $G_0$ summed over permutations of initial or final particle labels, and
\be
V=\frac{1}{2}(V_1+V_2+V_3),\hspace{1cm}
V_i(1'2'3,123) = v(j'k',jk) d^{-1}(i)\delta(i',i)      \eqn{V_sym}
\ee
[$(ijk)$ is a cyclic permutation of $(123)$] is the two-body potential with spectator particle $i$. Note that in this work we follow the literature and neglect three-body forces (the connected part of $V$). \eq{Gas} is a four-dimensional integral equation whose input two-body interactions originate from a model  second quantized Hamiltonian, e.g.
$H=\sum_1\w_1a^\dagger_1a_1+\sum_{1234}\bar 
v(1234)a^\dagger_1a^\dagger_2a_4a_3$.
For later convenience we also define the disconnected four-dimensional t matrices $T_i$ where
\be
T_i = V_i + \frac{1}{2}V_i G_0 T_i,\hspace{1cm}
T_i(1'2'3,123) = t(j'k',jk) d^{-1}(i)\delta(i',i)  .    \eqn{LS}
\ee


\subsection*{Three-body equal time quasi-potential}  

Being a four-dimensional integral equation, \eq{Gas} involves  relative times (or relative energies) as integration variables, and is therefore not very convenient for practical calculations. For this reason we implement a three-dimensional reduction of this equation. To do this we follow the current literature and effect this reduction by equating times in initial states, and separately, in final states.  Thus, in the double-degree of freedom formalism, our central quantity is the two-time Green function $\la G\ra$ which is obtained from the four-dimensional Green function ${G}$ by equating times as just described. In momentum space, $\la G\ra$ is given by
\begin{align}
i \la G \ra (E, \p'_1\p'_2 & \p'_3,\p_1\p_2\p_3)
=\int  \frac{d{p'_1}^{0}}{2\pi} \,\frac{d{p'_2}^{0}}{2\pi}\,\frac{d{p'_3}^{0}}{2\pi}
\,\frac{dp_1^{0}}{2\pi}\,\frac{dp_2^{0}}{2\pi}\,\frac{dp_3^{0}}{2\pi} \, G(p'_1p'_2p'_3,p_1p_2p_3)\nn[1mm]
&(2\pi)^2\, \delta({p'_1}^{0}+{p'_2}^{0}+{p'_3}^{0}-E)\, \delta(p_1^{0}+p_2^{0}+p_3^{0}-E).
\eqn{tt_main}
\end{align}
Similar expressions hold for one and two-body Green functions. 
In order to define the equal time quasi-potential, one needs to use the inverse of the equal-time disconnected Green function $\la G_0\ra$; however, in the many-body case this inverse may not exist. For example, the equal-time three-particle free Green function $\la G_0^f\ra$ projects onto the sub-space projected by\cite{tbp}
\be
{\cal N} = n(\p_1)n(\p_2)n(\p_3) +  \bn(\p_1)\bn(\p_2)\bn(\p_3);
\ee
as a result, $\la G_0^f\ra$ cannot be inverted in the full space of momenta. To get around this problem we introduce modified Green functions
\be
\tG_0 = \la G_0\ra + (ggg-{\cal N})\Delta ,\hspace{1cm}
\tG = \la G\ra + (ggg-{\cal N})\Delta^P .  \eqn{tG0}
\ee
For non-zero $\Delta$, $\tG_0$ is not singular and can be inverted. In \eq{tG0}, $ggg$ is a direct product of $g$'s (one for each particle).
It is important to note that $\tG_0$ is identical to $\la G_0\ra$ in the subspace projected by ${\cal N}$ (i.e.\ ${\cal N}ggg\tG_0 = {\cal N}ggg\la G_0\ra = \la G_0\ra$).
The operator $\Delta$ is required to be fully disconnected, but can otherwise be chosen according to one's own convenience. For the free case, one can write down the inverse of $\tG^f_0$ explicitly \cite{tbp}.
The three-dimensional quasi-potential $\tV$ is then defined to satisfy the equation
\be
\tG = \tG_0^P + \tG_0\tV \tG.   \eqn{Gasn}
\ee
Similar to \eq{V_sym}, the quasi-potential $\tV$ is expressible as
\be
\tV=\frac{1}{2}\left(\tV_1+\tV_2+\tV_3\right),\hspace{1cm}
\tV_i(1'2'3',123) = \tv'_i\, \delta(i',i) \eqn{tV}
\ee
where $\tV_i$ is a pair interaction with particle $i$ as spectator. It can be shown that
\be
\frac{\tV_i}{2} =  \tG_0^{-1}\left[ \la G_0 \frac{V_i}{2} G_0\ra + \la G_0 \frac{V_i}{2} G_0\frac{V_i}{2} G_0\ra
- \la G_0 \frac{V_i}{2} G_0\ra\tG_0^{-1}\la G_0 \frac{V_i}{2} G_0\ra + \ldots\right]  \tG_0^{-1} . \eqn{tVipert}
\ee
Even though \eq{tVipert} is an infinite series that is very difficult to sum,
what enters the three-body Faddeev equations is not $\tV_i$ but the pair-interaction t matrix, $\tT_i$, defined in terms of the quasi-potential $\tV_i$ by the Lippmann-Schwinger equation
\be
\tT_i = \tV_i + \frac{1}{2}\tV_i \tG_0 \tT_i ,\hspace{1cm} \tilde{T}_i(1'2'3',123) = \tilde{t}'_i\, \delta(i',i) . \eqn{tLS}
\ee
The task of constructing $\tT_i$ thus appears to be especially formidable. That is why most works on this subject keep only the linear term in the input two-body interaction \cite{schuck}. In our case, this would mean keeping only the first term of the series in \eq{tVipert}. This is a practical but unreliable way out of the problem. However, as we show below, there is another way of solving this problem which gives the exact and complete summation of \eq{tVipert} followed by an exact solution of the Lippmann-Schwinger equation, \eq{tLS}.

\subsection*{Exact three-body equal-time disconnected t matrix}

Defining the three and four-dimensional  disconnected Green functions $\tG^u_i$ and $G^u_i$ as
\be
\tG^u_i = \tG_0 + \frac{1}{2} \tG_0 \tV_i \tG^u_i,\hspace{1cm} G^u_i = G_0 + \frac{1}{2} G_0 V_i G^u_i,
\ee
it follows from \eq{LS} and \eq{tLS} that
\be
\tG^u_i = \tG_0 + \frac{1}{2} \tG_0 \tilde{T}_i \tG_0,\hspace{1cm} G^u_i =G_0 + \frac{1}{2} G_0 T_i G_0.  \eqn{tGi}
\ee
Taking equal times in the four-dimensional versions of \eq{tGi}, it is not hard to see that
\be
 \tG_0 \tT_i \tG_0 = \la G_0 T_i G_0\ra.   \eqn{main}
\ee
Thus, in contrast to the quasi-potential $\tV_i$ which is related to the four-dimensional potential $V_i$ in a very complicated way, the t matrix $\tT_i$ corresponding to the quasi-potential, defined by the {\em exact} solution of \eq{tLS}, is connected to the four-dimensional t matrix $T_i$ in a very simple way.
 
Using the disconnectedness structure given in \eq{LS} and \eq{tLS}, and repeating the above argument for two-particle Green functions, one can show that \eq{main} leads to the result
\be
 \tG_0 \, \tilde{t}'_i \, \tG_0  =   \tD_{0i} \tilde{t}_i \tD_{0i}\otimes\la d_i\ra
 \eqn{conv}
\ee
where $\tD_{0i}$ is the disconnected two-body propagator of particles $j$ and $k$ defined by analogy with $\tG_0$, and $\tilde{t}$ is the t matrix obtained from the two-body quasi-potential $\tv$. The symbol $\otimes$ denotes the convolution integral:
$
a\otimes b (E) \equiv  i/2\pi \int_{-\infty}^\infty a(E-z) b(z)\, dz.
$

\vspace*{-10ex}\ \\ 
\renewcommand{\refname}{}

\end{raggedright}

\begin{thebibliography}{0}
\setlength{\itemsep}{-1ex}
\bibitem{Beyer1} M. Beyer, W. Schadow, C. Kuhrts and G. R\"opke, Phys. Rev. C {\bf 60}, 034004 (1999).
\bibitem{Beyer2} M. Beyer, Few-Body Syst. Suppl.\ {\bf 10}, 179 (1999).
\bibitem{Beyer_quark}  M. Beyer, S. Mattiello, T. Frederico and H. J. Weber, Phys. Lett.  {\bf B521}, 33 (2001).
\bibitem{Mattiello}  S. Mattiello, M. Beyer, T. Frederico and H. J. Weber, Few-Body Syst., {\bf 31} 159 (2002); Few-Body Syst.\ Suppl.\ {\bf 14}, 379 (2003).
\bibitem{schuck} P. Schuck F. Villars and P. Ring, Nucl. Phys. {\bf A208}, 302.
(1973)
\bibitem{Beyer3} M. Beyer, G. R\"{o}pke and A. Sedrakian, Phys. Lett.\ {\bf B376}, 7 (1996).
\bibitem{Beyer4} M. Beyer, Nucl. Phys.\ {\bf A684}, 566c (2001).
\bibitem{BR} M. Beyer and G. R\"opke, Phys. Rev. C {\bf 56}, 2636 (1997).
\bibitem{walecka} A. L. Fetter and J. D. Walecka, {\em Quantum Theory of
Many-Particle Systems} (McGraw-Hill, New York, 1971).
\bibitem{dukelsky} J. Dukelsky, G. R\"{o}pke, and P. Schuck, Nucl. Phys. {\bf A628}, 17 (1998).
\bibitem{LeBellac} M. Le Bellac, {\em Thermal Field Theory} (Cambridge University Press, Cambridge, 1996).
\bibitem{smilga} A. V. Smilga, {\it Physics of Thermal QCD}, Phys. Rept. {\bf 291}, 1 (1997). 
\bibitem{umezawa} H. Umezawa, H. Matsumoto and M. Tachiki, {\em Thermo
Field Dynamics and Condensed States} (North-Holland, Amsterdam, 1982).
\bibitem{kbgaug} A. N. Kvinikhidze and B. Blankleider, Phys. Rev. C {\bf 60}, 044003
(1999) 
\bibitem{tbp} A. N. Kvinikhidze and B. Blankleider, in preparation.
\end{thebibliography}
\end{document}